\documentstyle[spie]{article} 
\input{psfig}   

\title{Searching for high-$z$ field ellipticals: successes and problems} 

\author{Andrea Cimatti\supit{a} 
\skiplinehalf 
\supit{a}Osservatorio Astrofisico di Arcetri, Largo Fermi 5, I-50125,Firenze,
Italy
}

\authorinfo{Further author information: (Send correspondence to A. Cimatti)\\
E-mail: cimatti@arcetri.astro.it}
 
\begin{document} 
  \maketitle 

\begin{abstract}
The most recent observational results on the search for high 
redshift field ellipticals are reviewed in the context of
galaxy formation scenarios. The perspectives for Large Binocular 
Telescope (LBT) observations are also discussed.
\end{abstract}

\section{INTRODUCTION}

The question on the formation of the present-day massive spheroidals 
is one of the most debated issues of galaxy evolution and it is 
strongly linked to the general problem of structure formation in the 
universe (see \cite{renz} for a recent review). 
In one scenario, massive spheroidals are formed at early
cosmological epochs (e.g. $z>3$) through the ``monolithic'' collapse 
of the whole gas mass\cite{egg,lar}. Such a formation would be characterized 
by an episode of intense star formation, followed by a passive 
evolution (or pure luminosity evolution, PLE) of the stellar population 
to nowadays.
In marked contrast, the hierarchical scenarios predict that massive 
spheroidals are the product of rather recent merging of pre-existing
disk galaxies taking place mostly at $z<1$\cite{k96,bau}. In hierarchical 
scenarios, fully assembled massive field spheroidals at $z>1$ are rare
objects\cite{k98}, and the spheroids of cluster ellipticals were assembled 
before those of field ellipticals\cite{bau}.
From an observational point of view, a direct way to test the above
scenarios is to search for massive field ellipticals at $z>1$ and to
compare their number with the model predictions (see the introduction
of \cite{schade} for a recent review on observational tests).

\section{HOW TO FIND $z>1$ ELLIPTICALS ?}

Since the near-IR light is a good tracer of the galaxy mass\cite{gava,k98}, 
$K$-band imaging provides an important possibility to
perform surveys aimed at selecting massive ellipticals at high-$z$.
A galaxy with a stellar mass of about $10^{11}$ M$_{\odot}$ is expected 
to have $18<K<20$ for $1<z<2$\cite{k98}, thus implying that moderately deep
$K$-band surveys can efficiently select massive galaxies. 

A first selection criterion to find ellipticals at $z>1$ is to
apply a color threshold to $K$-band selected galaxies. In the 
framework of passive evolution, such a threshold is set by
the colors expected for a galaxy at $z>1$ formed at a given 
$z_{f}$. For instance, according to the Bruzual \& Charlot
(1999) spectral synthesis models ($Z=Z_{\odot}$, Salpeter IMF), a very
red color of $R-K>5.3$ would allow to select $z\geq1$ passively evolving 
galaxies formed at $z_{f}>2$ ($H_0$=50 km s$^{-1}$ Mpc$^{-1}$,
$\Omega_0=0.1-1.0$), thus allowing to search for elliptical
candidates formed at early epochs. 
However, searches based on color selection criteria gave
discrepant results: some works showed that the number of such red 
galaxies is lower compared to the predictions of PLE\cite{zepf,fra,bar}, 
whereas others did not confirm such a deficit up to redshifts of about 
two\cite{tot,ben,bro,schade,scod}. 

In order to avoid the possible biases (e.g. star formation)
present in the color selection 
technique, another approach is to derive the fraction of ellipticals
by taking spectra of all the galaxies in $K$-selected samples 
irrespectively of colors\cite{cowie96,cohen,eisen,cim2001}
(see also {\tt http://www.arcetri.astro.it/$\sim$k20/}). This method 
allows to overcome the putative problem of ellipticals missed 
because bluer than the adopted color threshold due to a low 
level of residual star formation\cite{jim}. 

Finally, the third possibility to find $z>1$ ellipticals is to select 
galaxies according to their morphology and surface brightness profiles
(with or without an associated color selection criterion). This approach 
was adopted for example by \cite{schade} for $0.2<z<1.0$, and 
by \cite{fra,ben,bro,menan,treu,morio} for $z>1$. While the results seem 
to agree with no or little number density evolution for early type
galaxies at $z<1$, the analysis of the $z>1$ samples led again to 
discrepant results, thus making the question on $z>1$ ellipticals
even more controversial.

\section{RECENT IMAGING AND SPECTROSCOPY RESULTS}

After several small field surveys ($\sim$1-60 arcmin$^2$, see 
references in previous section) leading to discrepant results, 
the most recent success was provided by a wide field survey for 
extremely red objects (EROs)\cite{daddi1}. Such a survey (the widest 
so far: 700 arcmin$^2$ to $K<18.8$, with a sub-area of 447 arcmin$^2$ 
to $K<19.2$) provided a complete sample of about 400 objects with 
$R-K>5$ suitable for reliably constraining the number density of 
high-$z$ elliptical candidates.
The main results of such a survey are the detection of strong
angular clustering of EROs (an order of magnitude larger than that
of field galaxies; see Fig. 1-2), and the accurate estimate of the 
surface density of elliptical candidates at $z>1$. The observed 
clustering can explain the previous discrepant results on the 
surface density of $z>1$ ellipticals as due to strong field-to-field 
variations (the ``cosmic variance''), and it suggests 
that most EROs are ellipticals rather than dust reddened starbursts 
(see \cite{daddi1} for more details). Finally, even in the conservative 
case where up to 70\% of EROs are {\it not} ellipticals, the observed 
surface density (complemented by the results of \cite{tho}) is in good 
agreement with the predictions of PLE (Fig. 3), suggesting that most field 
ellipticals were fully assembled at least by $z=2.5$\cite{daddi2}. 
This result does not imply that the formation of massive spheroidals
occurred necessarily through a ``monolithic collapse'' scenario, but 
it simply constrains the epoch when the formation took place, and
it implies that, if ellipticals formed through merging, this occurred 
mostly at $z>2.5$.  

\begin{figure}
\begin{center}
\begin{tabular}{c}
\psfig{figure=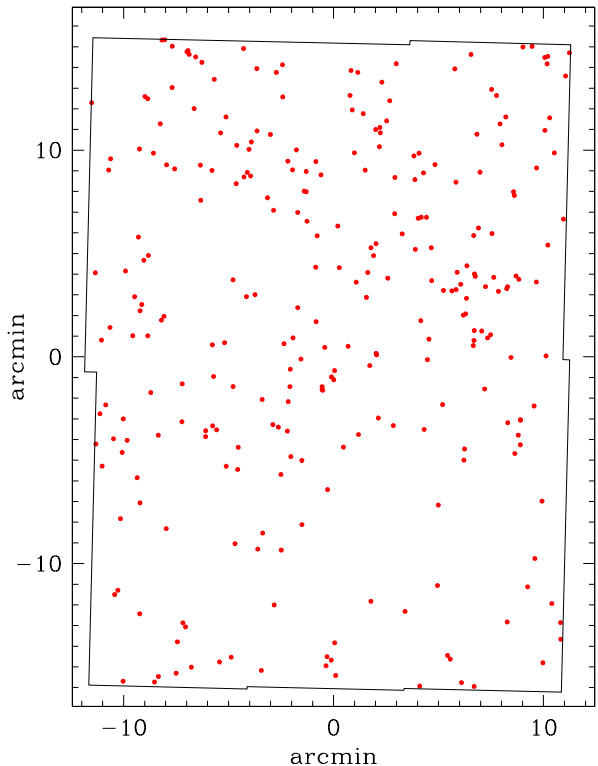,height=11cm}
\end{tabular}
\end{center}
\caption[fig1]{
The sky distribution of EROs with $R-K_s>5$ in the Daddi et al.
(2000) survey.
}
\end{figure}

\begin{figure}
\begin{center}
\begin{tabular}{c}
\psfig{figure=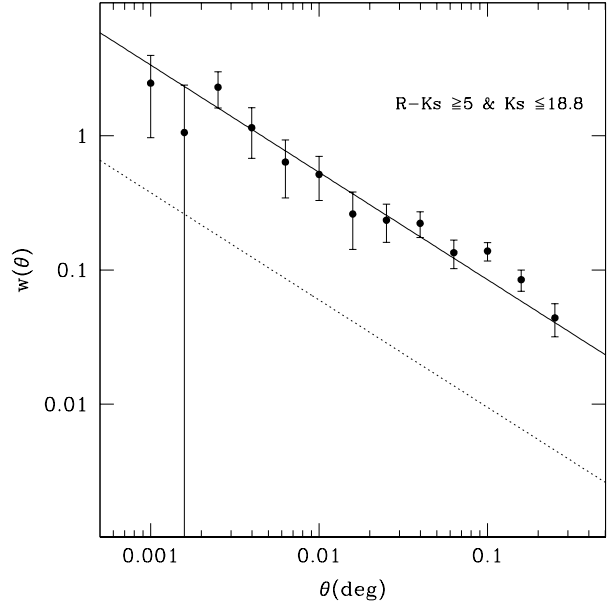,height=9cm}
\end{tabular}
\end{center}
\caption[fig2]{
The two-point angular correlation function of EROs with $R-K_s>5$ in 
the Daddi et al. (2000) survey. In comparison, the dashed line shows 
the lower clustering of the field $K$-selected galaxies.
}
\end{figure}

\begin{figure}
\begin{center}
\begin{tabular}{c}
\psfig{figure=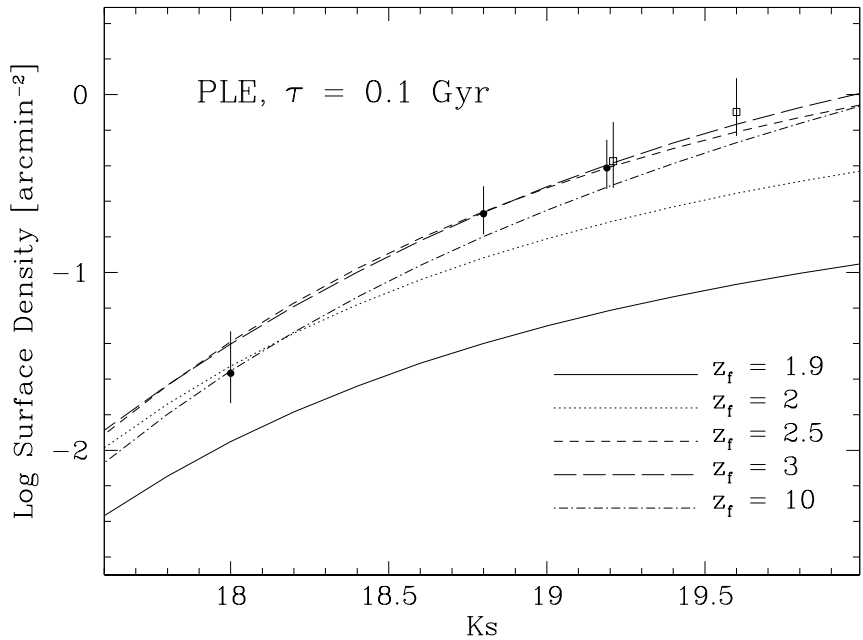,height=9cm,angle=-90}
\end{tabular}
\end{center}
\caption[fig3]{
The observed surface density of EROs\cite{daddi1,tho,daddi2} with 
$R-K_s>5.3$ (corresponding to $z>1$ selection)
compared to PLE models with a set of formation epochs (Bruzual \& 
Charlot 1999 models, $Z=Z_{\odot}$, Salpeter IMF, $\tau$=0.1 Gyr, 
Marzke et al. 1998 local luminosity
function of ellipticals, $z_{f}=1.9,2,2.5,3,10$).
}
\end{figure}

In addition to wide field imaging, the existence of galaxies with 
the colors expected in the case of passive evolution and with 
de Vaucouleurs $r^{1/4}$ surface brightness profiles consistent with being 
dynamically relaxed spheroidals at $z>1$ has significantly grown thanks 
to HST deep imaging (e.g \cite{ben,sti,morio}).

The spectroscopic confirmation of high-$z$ ellipticals is extremely
challenging because of their faintness both in the optical and
in the near-IR, and because of the few characteristic spectral
features present in their spectra: mainly the strong 4000~\AA~ 
continuum break, the weaker breaks at 2600~\AA, 2900~\AA~ and 3260~\AA, 
and a handful of absorptions detectable with the present-day largest 
telescopes and very long integration times (Fig. 4). Moreover, 
for $1.3<z<1.9$, the 4000~\AA~ break falls in a critical 
spectral region where the optical and near-IR spectrographs are
less efficient and the atmosphere severely hampers the observations.

Despite such difficulties, the Keck telescopes and the ESO VLT are 
confirming the existence of $z>1$ passively evolving ellipticals 
with old ages ($1-4$ Gyr) consistent with being formed at remote cosmological 
epochs\cite{spi,liu,soi,cim99,cim2001} (Fig. 4(a)), as well as 
ellipticals displaying a low level of star formation indicated by 
the possible detection of weak [OII]$\lambda$3727 emission (see Fig. 
4(b); see also\cite{schade}).

\begin{figure}
\begin{center}
\begin{tabular}{c}
\psfig{figure=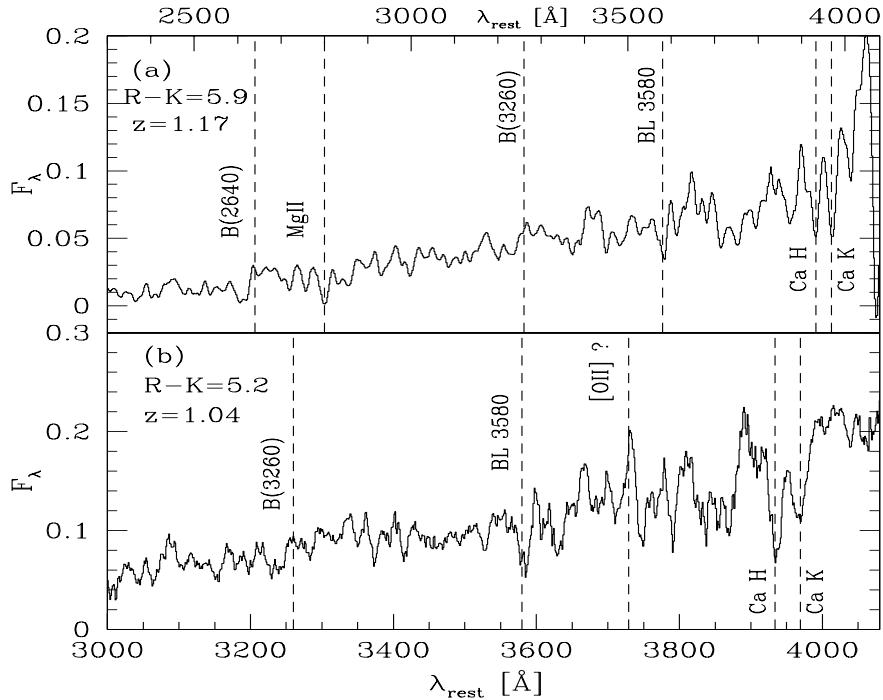,height=10cm,width=13cm}
\end{tabular}
\end{center}
\caption[fig4]{
ESO VLT-UT1 FORS1 spectra of $z>1$ ellipticals ({\it K20 survey}, 
Cimatti et al. 2001).
}
\end{figure}

\section{PROBLEMS}

The major problem affecting the statistical studies of $z>1$ 
ellipticals is the ``pollution'' of color-selected samples by a fraction 
of star forming galaxies reddened by strong dust 
extinction\cite{c98,d99,smail,and,gear}. The fraction of dusty EROs 
is currently unknown,
but its accurate estimate is crucial to infer a reliable surface density 
of high-$z$ ellipticals ``cleaned'' by the contamination of dusty systems.
This goal can be reached observing complete samples of EROs with 
optical+near-IR spectroscopy (when feasible), deep HST imaging and 
submm photometry. Although based on small and/or incomplete samples, 
recent submm and HST observations suggested that the dusty galaxies 
are probably segregated among the reddest EROs with $R-K>7$ or 
$I-K>6$\cite{c98,d99,smail,gear,morio}.

The other major problem in the identification of $z>1$ ellipticals is that 
a large fraction of EROs are beyond the spectroscopic limits 
of the present largest telescopes (e.g. $R>25$, $K>19-20$).
This strongly limits our ability to spectroscopically
confirm the nature of a high-$z$ elliptical candidate.

\section{PROSPECTS FOR THE LBT}

The Large Binocular Telescope will play an important role in the 
study of high-$z$ ellipticals. As a single-dish telescope, 
deep spectroscopy in the ranges 0.7-1.0$\mu$m and 0.9-1.8$\mu$m 
will be possible with the MODS and the LUCIFER spectrographs
respectively. This will allow us to enlarge the samples of 
spectroscopically identified high-$z$ ellipticals and, for instance, 
to infer their luminosity function and 3D clustering at $z>1$. As 
a diffraction limited telescope, the LBT will provide high 
angular resolution near-IR images that will be crucial in 
deriving the surface brightness profiles and, for instance, 
to extend the Kormendy relation and to attempt the Tolman test 
at $z>1$. Finally, in case of objects unfeasible with spectroscopy,
the combination of high resolution imaging and optical + near-IR 
photometry (possibly done with {\it special medium band} filters) 
will allow us first to confirm the spheroidal nature of the
elliptical candidates, and then to reliably estimate their
photometric redshifts. 

\acknowledgments
I am grateful to Alvio Renzini for reading this manuscript and for
the constructive comments, and to Shri Kulkarni for ``politely''
pointing out the ``high quality'' of the spectra.

\end{document}